\documentclass[twocolumn,showpacs,preprintnumbers,amsmath,amssymb]{revtex4}

\usepackage{graphicx}
\usepackage{dcolumn}
\usepackage{bm}

\begin{document}

\title{Dynamics of vortex glass phase in strongly type II superconductors}
\author{Qing-Hu Chen$^{1,2,3}$}
\address{$^{1}$ CSTCMP and Department of Physics, Zhejiang Normal University, Jinhua
321004, P. R.China\\
$^{2}$ Department of Physics, Zhejiang University, Hangzhou
310027, P. R.China\\
 $^{3}$ Computational Materials
Science Center, National Institute for Materials Science, Tsukuba
305-0047, Japan}

\date{\today}

\begin{abstract}
Dynamics of vortices in strongly type-II superconductors with strong
disorder is investigated within the  frustrated three-dimensional XY
model. For two typical models in [Phys. Rev. Lett. {\bf 91}, 077002
(2003)] and [Phys. Rev. B {\bf 68}, 220502(R) (2003)],  a strong
evidence for the finite temperature vortex glass transition in the
unscreened limit is provided by performing large-scale dynamical
simulations. The obtained correlation length exponents and the
dynamic exponents in both models are different from  each other and
from those in the three-dimensional gauge glass model. In addition,
a genuine continuous depinning transition is observed at zero
temperature for both models. A scaling analysis for the thermal
rounding of the depinning transition shows a non-Arrhenius type
creep motion in the vortex glass phase, contrarily to the recent
studies.
\end{abstract}


\maketitle

\section{Introduction}

The  application of superconductors crucially depends on the high
electric current density without dissipation. However, the
resistivity would always be nonzero even in the presence of pining
centers. This conventional picture has been changed with the
discovery of high-$T_c$ superconductors\cite{Review} and the
progress in random field systems\cite{Young}. Similar to the
spin-glass system, Fisher et al suggested that, for strong disorder,
the system freezes into a genuine thermodynamic amorphous vortex
glass (VG) phase  with some kind of glassy long-range
orders\cite{theory,FFH}. The VG phase in strongly type II
superconductors has attracted considerable attentions both
experimentally and theoretically\cite{Nattermann} during the past
more than one decades. It is of practical significance that the VG
phase is a true superconducting state with a vanishing linear
resistivity by diverging energy barriers. On the fundamental side,
it is closely related to an important class of phenomena in
condensed matter physics, such as spin glasses, random field
systems\cite{Young}, and charge-density waves in
solids\cite{Gruner}.

The evidences to support the existence of a VG phase have been
reported in many experiments by means of the dynamic scaling of the
measured current-voltage (IV) data\cite{exp}.  However,  Strachan et
al. have shown that  a perfect collapse of the IV data is not the
sufficient evidence for a VG transition \cite{Strchan}, since the
critical temperature and the scaling exponents are not uniquely
determined by this dynamic scaling.

Theoretically,  the XY gauge glass
model\cite{Reger,Olson,Katzgraber,Kosterlitz} has been extensively
employed to study the VG transition.  The values for the critical
exponents are similar to those obtained in  some experiments
\cite{exp}. However, lacking some of properties and symmetries due
to the absence of net magnetic fields, it is questioned to be a
model of disordered superconductors in an applied
filed\cite{Vestergren,Olsson,Kawamura,Olsson1,Lidmar}. Some
realistic models have  then been  proposed recently, but the
conclusions were quite contradictory. Continuous finite temperature
VG transitions have been given by most  models with various critical
exponents. It was also observed that the VG phase disappears if the
screening of the vortex interaction due to the gauge field
fluctuation is included\cite{Rieger}. The simulation of the
London-Langevin model suggested no VG phase\cite{against}.

Among all vortex models, the disordered three-dimensional (3D) XY
model with net magnetic fields has provided both equilibrium and
dynamical vortex phase diagrams in Type-II superconductors with weak
disorder \cite{Nonomura,chen2}. The low field (weak disorder) low
temperature phase is in general regarded as a dislocation-free Bragg
glass with  a quasi-long-range order\cite{Nonomura}, which was
observed directly in a neutron experiment\cite{Klein}. Several
dynamical simulations on the vortex matter with rather low fields
for weak disorder have been performed in this
model\cite{chen2,Olsson2,Hern}. By a anisotropic frustrated 3D XY
model with strong disorder in the coupling constants, Olsson
\cite{Olsson} provided evidence for the VG transition in the
unscreened limit. The correlation length exponent $\nu=1.5\pm 0.3$
was obtained, consistent with the 3D gauge glass universality.
Within an isotropic model with different choice of strong
random-coupling distribution,   Kawamura \cite{Kawamura} reported
similar results for the  VG transition independently. Although the
obtained value $\nu=1.2\pm 0.3$ is slightly smaller, which within
the error bar also suggests a common universality with the 3D gauge
glass model. However, it was found later that a convincing scaling
collapse for helicity modulus could not be achieved in Kawamura's
model\cite{Olsson1}, possibly due to the small effective randomness
in the  small system accessed. Furthermore, to the best of our
knowledge, the dynamical study in  the frustrated 3D XY model with
strong disorder is so far lacking, which is however more relevant to
experiments in the context of VG transitions.

In this paper, based on resistively-shunted-junction dynamics, we
perform large scale dynamical simulations  on the frustrated 3D XY
models for two typical sets of parameters in Refs.
\cite{Olsson,Kawamura}. The glass transition temperatures  and the
critical exponents are estimated based on the simulated IV data. The
depinning transition at zero-temperature and creep motion far below
the glass transition temperature are also studied.  The rest of the
paper is organized as follows. Sec.II describes the models and
dynamic method. In Sec. III and Sec. IV,  the main results for the
VG transition, the depining and creep motion of vortices are
presented, and some discussions are also carried out. Finally, a
short summary is given in the last section.

\section{Model}

The frustrated 3D XY model on a simple cubic lattice is given by
\cite{Olsson,Kawamura}
\begin{equation}
H=-\sum_{\langle ij\rangle }J_{ij}\cos (\phi _{i}-\phi
_{j}-A_{ij}), \label{Hamil}
\end{equation}
where $\phi _{i}$ specifies the phase of the superconducting order
parameter on site $i$, $A_{ij}=(2\pi /\Phi _{0})\int_{i}^{j}{\bf
A\cdot dl}$ with $ {\bf A}$ the magnetic vector potential of a field
${\bf B}=\nabla \times {\bf A}$ along the $z$ axis, $J_{ij}$
represents the random coupling distribution. The average number of
vortex lines per plaquette is denoted by $f=l^{2}B/\Phi _{0}$, where
$l$ is the grid spacing in the $xy$ plane and $\Phi _{0}$ is the
flux quantum. We choose two typical sets of parameters used by
Olsson\cite{Olsson} and Kawamura\cite{Kawamura}. For convenience,
the models with these parameters are called Models I and II
respectively. In Model I,  the random pinning potential is
introduced in the coupling strength in the $xy$ plane
$J_{ij}=J_{0}(1+p\epsilon _{ij})$, where $\epsilon _{ij}$'s are
independently Gaussian distributed with zero mean and unit variance,
$p$ represents the pinning strength. The coupling between the $xy$
planes is $J_{z}=J_{0}/\Gamma ^{2},$ ($\Gamma $ is the anisotropy
constant). We typically choose $p=0.4$ which models strong pinning
strength, $1/\Gamma ^{2}=1/40$, and $f=1/5$. Simulations of Model I
are performed with system size $L_{xy}=100, L_{z}=60$ satisfying
$L_{xy}/L_{z}=5/3$, much too larger than those in Ref.
\cite{Olsson}. In Model II, the quenched randomness is put in the
coupling constant $J_{ij}$ in all directions, which is distributed
uniformly on the interval $[0, 2 J_{0}]$. The filling factor is
chosen to be $f=1/4$. The present simulations of Model II are
performed with the system size $L=64$ for all directions,
considerably larger than those in Ref. \cite{Kawamura}.

The Resistivity-Shunted-Junction dynamics is incorporated in
simulations, which can be described as
\begin{equation}
{\frac{\sigma \hbar }{2e}}\sum_{j}(\dot{\phi _{i}}-\dot{\phi _{j}})=-{\frac{%
\partial H}{\partial \phi _{i}}}+J_{{\rm ext},i}-\sum_{j}\eta _{ij},
\end{equation}
where $J_{{\rm ext},i}$ is the external current which vanishes
except for the boundary sites. The $\eta _{ij}$ is the thermal
noise current with zero mean and a correlator $\langle \eta
_{ij}(t)\eta _{ij}(t^{\prime })\rangle =2\sigma k_{B}T\delta
(t-t^{\prime })$. In the following, the units are taken of
$2e=J_{0}=\hbar =\sigma =k_{B}=1$.

In the present simulation, a uniform external current $I_{x}$
along $x$-direction is fed into the system, analogous to
exeriments\cite{exp}. The fluctuating twist boundary condition
\cite{chen1} is applied in the $xy$ plane to maintain the current,
and the periodic boundary condition is employed in the $z$ axis.
In the $xy$ plane, the supercurrent between sites $i$ and $j$ is
now given by $\ J_{i\rightarrow j}^{(s)}=J_{ij}\sin (\theta
_{i}-\theta _{j}-A_{ij}-{\bf r} _{ij}\cdot {\bf \Delta })$, with
${\bf \Delta }=(\Delta _{x},\Delta _{y})$ the fluctuating twist
variable and $\theta _{i}=\phi _{i}+{\bf r}_{i}\cdot {\bf \Delta
}$. The new phase angle $\theta _{i}$ is periodic in both $x$- and
$y$-directions. Dynamics of ${\bf \Delta }_{\alpha }$ can be then
written as
\begin{equation}
\dot{\Delta}_{\alpha }={\frac{1}{L^{3}}}\sum_{<ij>_{\alpha
}}[J_{i\rightarrow j}^{(s)}+\eta _{ij}]-I_{\alpha }, \alpha =x,y.
\label{delta-dot}
\end{equation}
The voltage drop is $V=-L{\dot{\Delta}_{x}}$.

The above equations can be solved efficiently by a pseudo-spectral
algorithm \cite{chen2} due to the periodicity of phases in all
directions. The time stepping is done using a second-order
Runge-Kutta scheme with $\Delta t=0.05$. The equilibration of the
simulation should be ensured before the measurement. So most of our
runs are typically $(4-8)\times 10^{7}$ time steps and the latter
half time steps are for the measurements.  The detailed procedure in
the simulations was described in Ref. \cite{chen2}. Our results are
based on one realization of disorder. The  present system size is
much too larger than those reported in literature, it is expected to
exist a good self-averaging effect.  We have done two additional
simulations with different realizations of disorder for further
confirmations, and indeed observed quantitatively  the  same
behavior. In addition, it is practically difficult to perform any
serious disorder averaging for such a rather large system. Actually,
the results from dynamic simulations on 3D XY model  in the recent
literature were also for a single disorder
realization\cite{chen2,Olsson2,Hern}, mainly due to the large system
simulated. For the data points presented in the following figures,
the statistical errors are smaller or comparable to the symbol
sizes.

\section{VG phase transitions }

\begin{figure}[tbp]
\centering
\includegraphics[width=6cm]{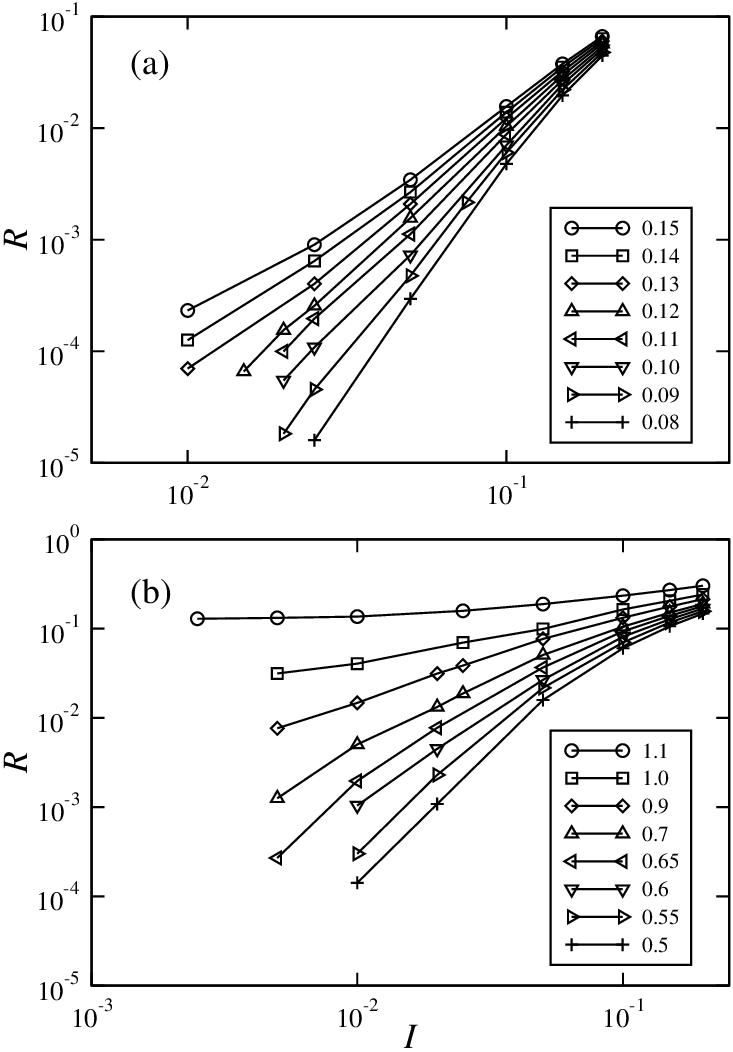}
\caption{Current-resistivity curves  at various temperatures for (a)
Model I and (b) Model II} \label{Fig1}
\end{figure}

First, we study the  VG phase transition in these two models. In
Model I,  the VG transition temperature $T_g$ is estimated to be
$0.123 \pm 0.008 $ in equilibrium simulations\cite{Olsson}. The IV
characteristics is simulated at  temperatures ranged from  $ 0.08$
to $0.15$, which must covers possible $T_g$. In the equilibrium
simulations of Model II, Kawamura\cite{Kawamura} obtained
$T_{g}=0.81$ by performing the finite-scaling analysis of the Binder
ratio and the mean-square current. The similar simulations on Model
II are performed at the temperatures ranged from  $0.5$ to $ 1.1$,
which also covers the possible $T_g$. In simulations on  both
models, we try to probe the system at currents as low as possible
for each temperature. Figs. 1(a) and 1(b) present the resistivity
$R=V/I$ as a function of current $I$ at various temperatures for
Models I and II, respectively.  It is clear that, at lower
temperatures, $R$ tends to zero as the current decreases, which
follows that there is a true superconducting phase with zero linear
resistivity. While $R$ tends to a finite value at higher
temperatures, corresponding to an Ohmic resistivity in the vortex
liquid. These observations provide a evidence of the existence of
the VG phase in both models.

Assuming  that the vortex glass transition is continuous and
characterized by the divergence of the characteristic length and
time scales $t\sim\xi^{z}$ (z is the dynamic exponent) , Fisher,
Fisher, and Huse \cite {FFH} proposed the following dynamic
scaling ansatz to analyze the glass transition from a vortex
liquid with ohmic resistance to a superconducting glass state,
\begin{equation}
TR\xi^{1-z}=\Psi_{\pm }(I\xi^{2}/T).  \label{ffh}
\end{equation}
where $\xi\propto\mid T/T_g-1\mid^{-\nu}$  is the correlation length
which diverges at the transition. $\Psi(x)$ is a scaling function
with the + and - signs corresponding to $T>T_g$ and $T<T_g$. Eq.
(\ref{ffh}) was often used to scale measured IV data
experimentally\cite{exp}.

Right at $T_g$, the $RI$ curve should show a power law behavior $ R
\propto I^{-\alpha}$ where $\alpha=(z-1)/2$, which provides a
convexity-concavity criterion to identify the VG transition
temperature as well as the dynamic exponent $z$. As shown in Figs.
1(a) for Model I that the value of $T_g$ is in between
$(0.12-0.13)$, because in the low current regime the $RI$ curves
show convexity below $T=0.12$ and  concavity  above $T=0.13$. The
$RI$ curves  at other temperatures within $(0.12,0.13 )$ can be
obtained by  interpolations. The temperature at which the $RI$ curve
most close to the power law behavior is regarded as $T_{g}$ and the
$RI$ power law exponent at $T_g$ gives the dynamic exponent $z$. The
error bars are estimated by obvious deviation from the power law
behavior. In this way, for Model I, we obtain $T_{g}=0.124\pm 0.002$
and $z=5.8\pm 0.3$. The value of $T_g$ is consistent with that in
equilibrium simulations\cite{Olsson}. By the similar method, for
Model II, we get $T_{g}=0.81\pm 0.01$ , $ z=2.5\pm 0.2$.
Interestingly, the present value of $T_g$ in Model II agrees well
with that in equilibrium Monte Carlo simulations by
Kawamura\cite{Kawamura}.

\begin{figure}[tbp]
\centering
\includegraphics[width=6cm]{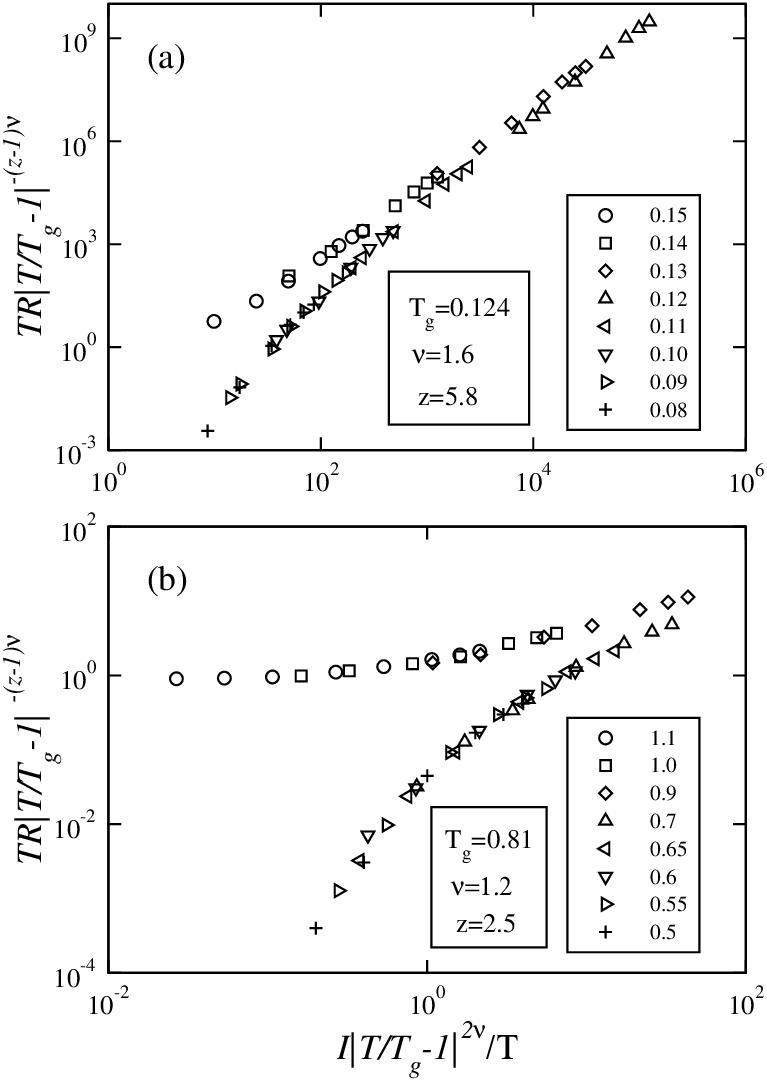}
\caption{Dynamic scaling of IV data  at various temperatures for (a)
Model I and (b) Model II} \label{Fig2}
\end{figure}

Once $T_g$ and $z$ have been estimated, we can examine the IV data
at different temperatures by the dynamical scaling. Fig. 2(a) shows
that the data collapses well according to Eq. (\ref{ffh}) if using
the correlation length  exponent $\nu =1.6\pm 0.1$. The error bar is
estimated by tuning the value of $\nu$ until the collapse becomes
poor evidently. The value of $\nu$ is  very close to $\nu = 1.5\pm
0.3$ obtained in Ref. \cite{Olsson} through equilibrium Monte Carlo
simulations of Model I. Also as indicated in Fig. 2(b) that, using
$\nu =1.2\pm 0.1$, an excellent collapse according to Eq. (4) is
achieved. The value  of $\nu$ also agrees quite well with $\nu =
1.2\pm 0.3$  estimated in an equilibrium Monte Carlo simulations of
Model II\cite{Kawamura}. Interestingly, although the values of $\nu$
in both Models  lie in the range $[1.0-2.0]$ usually observed
experimentally \cite{exp}, they  are close to but different from
each other. Since the present two models involve different
symmetries (anisotropy) and different disorders included, in our
opinion, it is not unlikely that they represent different
universality classes.

It should be mentioned that the present analysis method for the VG
phase transition  is not essentially inconsistent with that
described in Ref. \cite{Strchan}. We also think that only the
perfect collapse of the IV data is not a sufficient evidence for a
VG transition, so we use the convexity-concavity criterion to
identify $T_g$ and determine $z$ before performing the dynamic
scaling.

In equilibrium Monte Carlo simulations of  Model II,  some
quantities failed to provide good scaling\cite{Olsson1}. The
helicity modulus was used in the finite size scaling analysis of the
VG phase transitions in both models\cite{Olsson,Olsson1}, a nice
scaling is obtained in Model I, but scaling fails applied to Model
II for data in system sizes $L \le 20$. The collapse of the
transverse helicity modulus with poor quality gives $T_{g}=0.63$,
$\nu = 1.5$, which differed significantly from those in Ref.
\cite{Kawamura}. More seriously, it was impossible to collapse the
data for the parallel helicity modulus. It has been
observed\cite{Nonomura} that the correct behavior required a great
flexibility of the field induced vortex lines, which could be
obtained either with a very large size  or with weak interplane
coupling along the field direction. For the isotropic system in
Model II, the possible way to get a convincing scaling collapse of
some quantities  is to enlarge the systems. In the present large
scale dynamical simulations, a excellent collapse of the IV data in
the dynamic scaling is indeed achieved.

The exponents $\nu$  in the present two models  are close to
$\nu=1.39\pm 0.20$ evaluted by Olson and  Young\cite{Olson} ,  but
different from the recent more accurate result $\nu=1.39\pm 0.05$
obtained by Katzgraber and  Campbell \cite{Katzgraber}  in the 3D
gauge glass model, suggesting that they are not in the same
universality class. It follows that the difference in the quenched
randomness and the introduction of net  fields may change the static
critical properties of the VG transitions.

The dynamic exponents $z$ in these two models are found to be quite
different. The exponent $z$ in Model I is high, in the range
$[4.0-6.0]$ usually measured in experiments\cite{exp},  in Model II
is however considerably low. Note that small values of the exponent
$z$ were also reported\cite{Gammel}. In addition, both exponents $z$
in Models I and II can not fall even within the error bar of that in
the 3D gauge glass model, which were estimated  to be $z=4.2\pm 0.6$
in Ref. \cite{Olson} and  $z=4.7\pm 0.1$ in Ref. \cite{Katzgraber},
although the exponent $z$ in Model I seems to be more close.  It is
possible that the disorder in the coupling constant along the filed
direction in Model II reduces the effective pinning strength
\cite{Olsson1}, leading to a small  IV power law exponent at the VG
transition. It is not expected that enlarging the system size
further along the field direction would change the dynamic exponent
$z$ essentially. Nevertheless, the reason for the small value of $z$
in Model II is not fully understood at the present stage, the
further investigation is clearly called for.

\section{Depining and creep}

\begin{figure}[tbp]
\centering
\includegraphics[width=7cm]{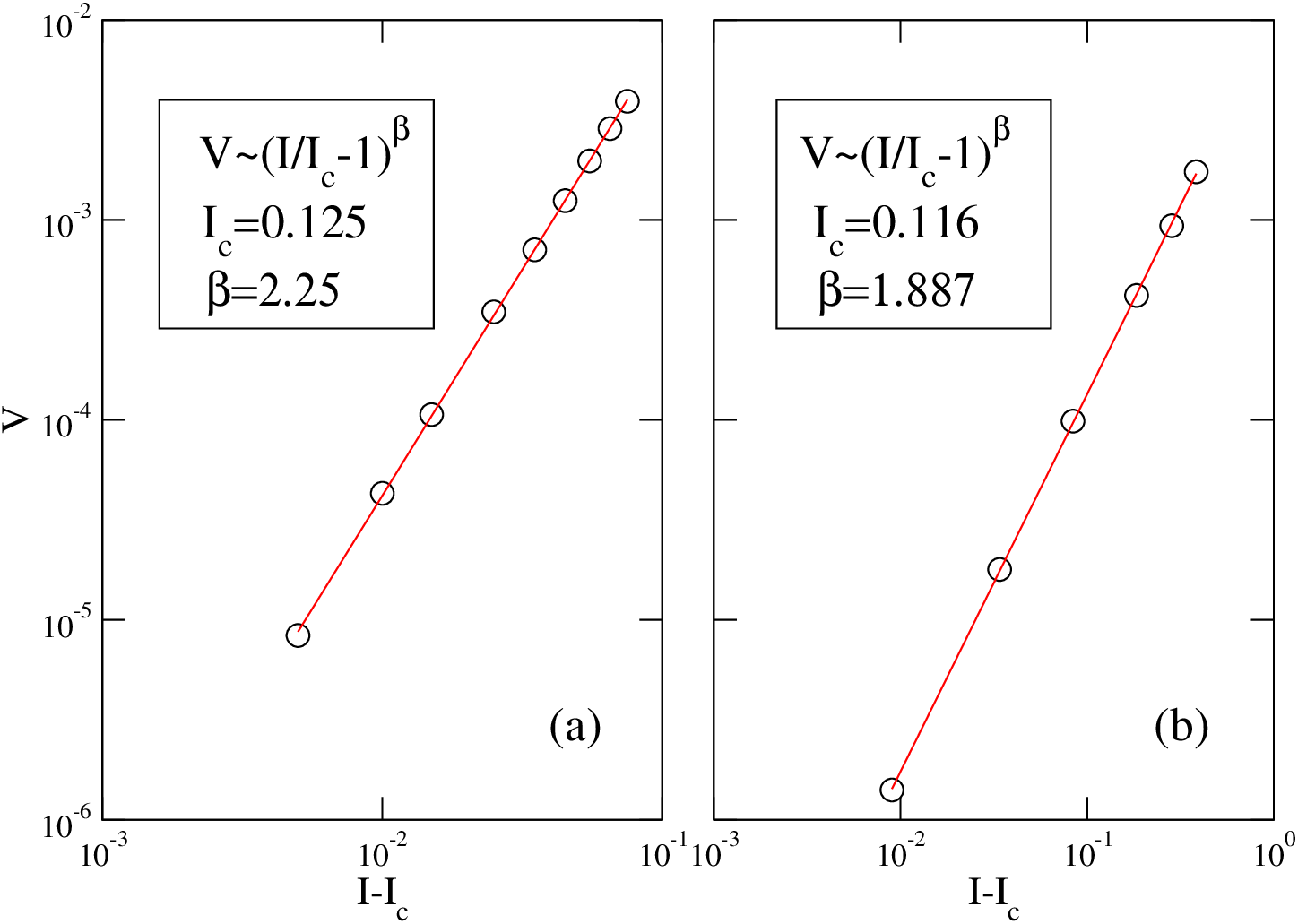}
\caption{Log-log plots of $IV$ data at zero-temperatures for (a)
Model I and  (b) Model II.} \label{Fig3}
\end{figure}

With the VG phase in hand,  we then study the depining and creep
motion of the vortices in this phase for both models.  To study the
depinning transition at zero temperature, we start from high
currents with random initial phase configurations. The current  is
then lowered step by step. The steady-state phase configurations
obtained at higher currents are chosen to be the initial phase
configurations of the lower currents in the next step. It becomes
more difficult to measure the voltage with the lower currents. In
the vicinity of the critical current, a huge amount of the computer
time is consumed to get accurate results. Fig. \ref{Fig3} exhibits
the $IV$ characteristics at $T=0$ for both models. Interestingly, we
observe continuous depinning transitions with unique depinning
currents\cite{Fisher}, which can be described as $V\propto
(I-I_{c})^\beta$ with $I_{c}=0.125\pm 0.001$, $\beta=2.25\pm 0.02$
for Model I and $I_{c}=0.116\pm 0.002$, $\beta=1.887\pm 0.01$ for
Model II. Note that the depinning exponents for both models are
greater than $1$, consistent with the mean  field studies on
charge-density wave models\cite{Fisher}. The depinning exponent is
model dependent, possibly due to the different realizations of  the
disorder.

At low temperatures,  the $IV$ curves are rounded near the
zero-temperature critical current due to thermal fluctuations. An
obvious crossover between the depinning and creep motion can be
observed around $I_{c}$ for both models at the lowest accessible
temperatures. In order to address the thermal rounding of the
depinning transition, Fisher\cite{Fisher} first suggested to map
this system to the ferromagnet  in fields where the second-order
phase transitions occur. This mapping  was latter extended to the
random-field Ising model\cite{Roters} and flux lines in type-II
superconductors\cite{luo}. If the voltage is identified as the order
parameter, the current and temperature are identified as the inverse
temperature and the field in the ferromagnetic system respectively,
analogous to the second-order phase transitions, a scaling relation
among the voltage, current, and temperature in the present model
should satisfy the form
\begin{equation}~\label{scaling1}
V(T,I)=T^{1/\delta}S[T^{-1/\beta\delta}(1-I_{c}/I)],
\end{equation}
where $S(x)$ is a scaling function.

\begin{figure}[tbp]
\centering
\includegraphics[width=7cm]{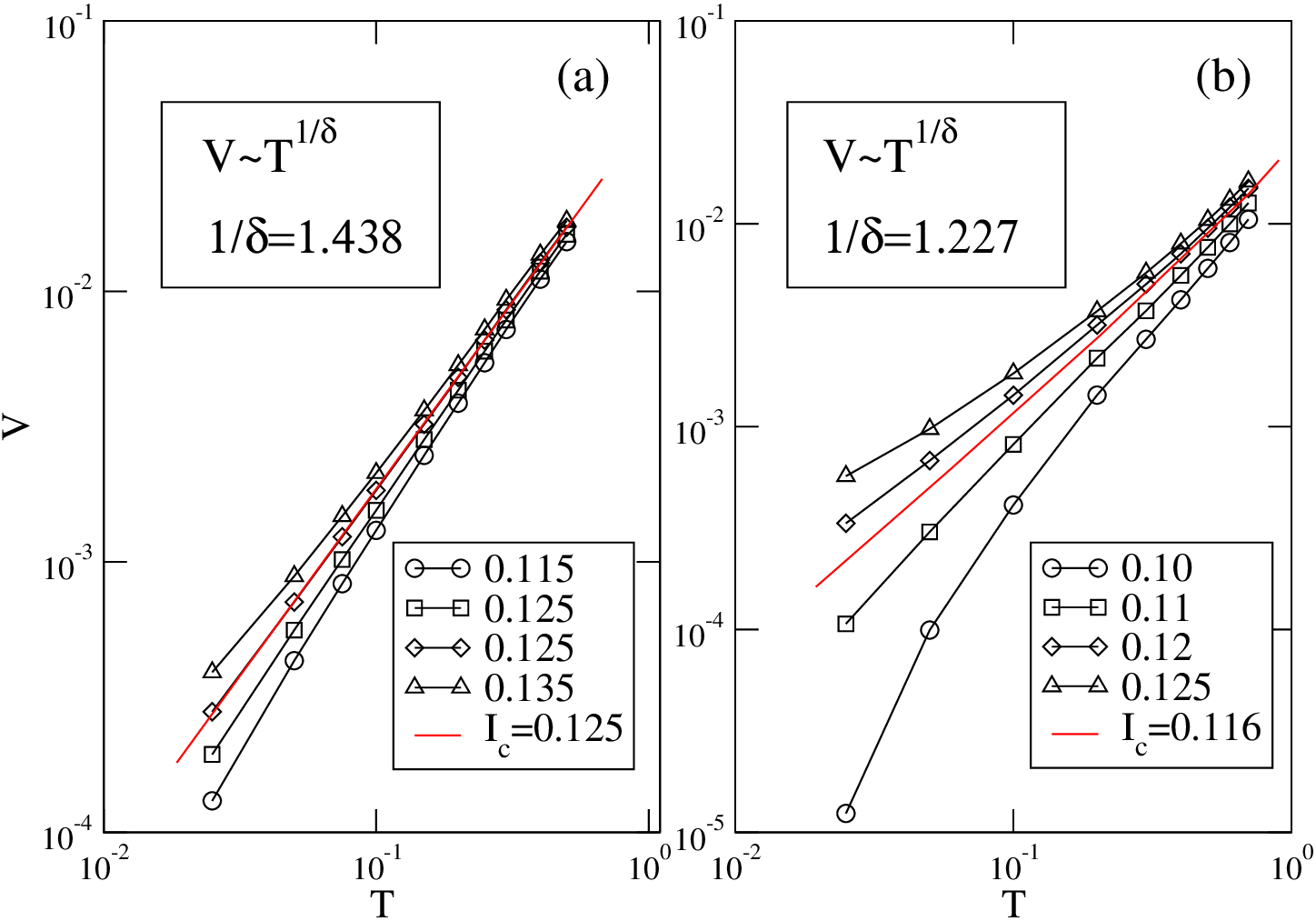}
\caption{Log-log plots of $V-T$ at three currents around $I_{c0}$
for (a) Model I and  (b) Model II.} \label{Fig4}
\end{figure}

It is implied that right at $I=I_{c}$ the voltage shows a power-law
behavior $V(T,I=I_{c})\propto T^{1/\delta}$ and the critical
exponent $1/\delta$ can be determined. The log-log $V-T$ curves are
plotted in Fig. \ref{Fig4} (a) and (b) at three currents for Models
I and II. In Fig. \ref{Fig4}(a), we can see that the critical
current is between $0.115$ and $0.135$. The values of voltage at
other currents within $(0.115,0.135 )$ can be evaluated by quadratic
interpolations. The deviation of voltage from the power law is
calculated as the square deviations. The current at which the square
deviation is minimum is defined as the critical current
$I_{c}=0.125\pm 0.02$, consistent with those obtained at zero
temperature. The temperature dependence of voltage at the critical
current is also plotted in Fig. \ref{Fig4}(a). The slope of this
curve yields $1/\delta=1.438\pm 0.004$. The similar analysis in Fig.
4(b)  yields $I_{c}=0.116\pm 0.02$ for Model II,  consistent with
that extracted from the zero-temperature simulation. The exponent
$1/\delta=1.227 \pm 0.003$ is achieved by  fitting the $V-T$ curve
in the low temperature regime at the critical current.

\begin{figure}[tbp]
\centering
\includegraphics[width=6cm]{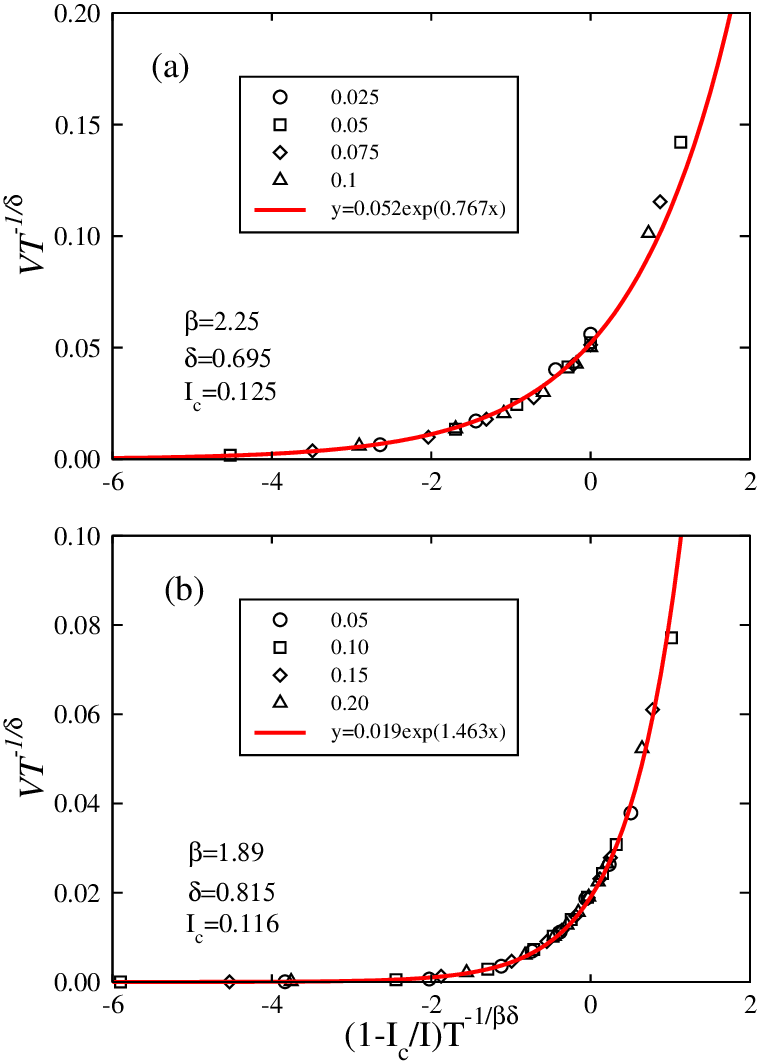}
\caption{Scaling plot  of the IV data at  various temperatures
below $T_g$ for (a) Model I and  (b) Model II.} \label{Fig5}
\end{figure}

With the critical exponent $\delta$ and the critical current
$I_{c}$, we can adjust the depinning exponent $\beta$ to achieve the
best data collapse according to the scaling relation Eq.
(\ref{scaling1}) for $I\le I_c$. In Fig. \ref{Fig5}  (a) and (b), a
perfect collapse of the IV data at various temperatures below $T_g$
is shown with  $\beta=2.25 \pm 0.01$ for Model I and $1.89 \pm 0.01
$ for Model II. The values of $\beta$ estimated from low temperature
creep motion are in excellent agreement with those derived at $T=0$
depinning transition. Moreover, the scaling function with the form
$V \propto   T^{1/\delta }\exp [A (1-\frac{I_c}I)/T^{\beta\delta}]$
is derived in the creep regime for both models, which are also
demonstrated  in the legends of Figs. \ref{Fig5} (a) and (b). Note
that the product of the two exponents $\beta\delta$ describes the
temperature dependence of the creeping law. Interestingly,
$\beta\delta \simeq 1.56$ for Model I and $\beta\delta \simeq 1.54$
for Model II are obtained, both deviate from unity, demonstrating
that the creep law is a non-Arrhenius type. The values of
$\beta\delta$ in both models are close to $3/2$, which may motivate
a further analytical work. In our opinion,  it is not a coincidence
that they are   in the same universality class in the depinning
transition.

In a recent study of the depinning and creep motion of the flux line
system in type-II superconductors\cite{luo}, by simulations of
overdamped London-Langevin model,  Luo and Hu observed an Arrhenius
law for the creep motion with a linearly suppressed energy barrier
for strong pinning \cite{luo}, inconsistent with the present study
for strong disorder. It is worth noting that, in the London-Langevin
model, the stable VG phase with the freezing of disordered vortex
matter is not found\cite{against} and instead the vortices freeze
like a window glass, called vortex molasses scenario. In the
framework of the frustrated 3D XY model, the existence of a stable
VG phase is well established in the unscreened limit in the present
dynamical simulations, as well as in previous equilibrium Monte
Carlo simulations\cite{Olsson,Kawamura,Olsson1}. In the real
strongly type-II superconductors, the screening induced rounding of
the sharp VG transition is only a weak effect, and only visible at
temperatures very close to $T_g$. The present good scaling behavior
in the creep motion is just observed far below $T_g$. We believe
that the different nature of the phase in the  London-Langevin model
with strong pinning \cite{luo} and the VG phase in the present two
models  is the possible reason for the discrepancy. In addition, the
Anderson-Kim creep law \cite{Anderson} is realized in the
London-Langevin model with strong pinning \cite{luo}, which may
suggest that it is applicable to strong flux pinning in the
conventional low $T_c$ superconductors rather than a VG phase with
random point pins.

The non-Arrhenius type creep behaviors have been also observed in
charge-density waves with the mean-field result
$\beta\delta=2/3$\cite{Middleton}, the 3D random-field Ising model
with $\beta\delta\approx 3/2$\cite{Roters}, $(1+1)$ elastic
interface with $\beta\delta\approx 2$\cite{Bustingorry}, and the
flux line system in type-II superconductors for weak pinning in a
Bragg glass phase with $\beta\delta\approx 3/2$\cite{luo}. It is
surprising  to note that the present combined exponent
$\beta\delta\approx 3/2$  in the frustrated 3D XY model for  strong
disorder is close to that in the 3D London-Langevin model  for
weak-pinning \cite{luo}. In the London-Langevin model of a fixed
number of interacting particles, the vortex loop between the planes
perpendicular to the field  is absolutely excluded, while in the
present 3D XY models, the vortex loops between the planes can be
induced by both thermal activations and the quenched disorder. So we
argue that the  disorder strengths in these two different kinds of
models  are hard to compare. Interestingly, the combined depinning
exponent $\beta \delta\approx 3/2$  was also observed in the
depinning of domain walls in the 3D random-field Ising model
\cite{Roters}, possibly suggesting a universal rule in high
dimensional elastic systems. Whereas the present results are
different from recent results for (1+1) elastic interfaces in a
disorder medium\cite {Bustingorry}, possibly owing to the
two-dimensional  nature in the latter. Further work is needed in
order to clarify these observations.

Note that the depinning of Bragg glass phase has been studied
recently by  Olsson  using essentially the same model as Model I
with a rather weak field $f=1/45$ \cite{Olsson2}. The IV
characteristics showed a unexpected behavior with a critical current
that separates the creep region with an immeasurably low voltage at
$I<I_c$ from a region with $V \propto (I-I_{c})$.  This behavior is
not observed in the present two models, possibly owing to the strong
disorder and high fields. Their study together with the present one
constitute a complementary picture for the depinning  in the
disordered 3D XY model with net fields.

\section{Summary}

We have performed large scale dynamical simulations on the
frustrated 3D XY models for strong disorder with  two typical sets
of parameters used in recent literature, within the
resistively-shunted-junction dynamics.  We first use the
convexity-concavity criterion to identify $T_g$ and determine $z$,
then perform the dynamic scaling on the simulated IV data. Adjusting
the single parameter $\nu$, a perfect collapse of  IV data is
achieved for both cases, providing a evidence of the VG transition
in the unscreened limit convincingly. Although the obtained
correlation length exponents agree with the previous ones within
error bars in equilibrium simulations, they are close to but
different from each other, suggesting different universality class.
New dynamic exponents are found to be parameter dependent. Both the
static and dynamic exponents are different from the recent accurate
results in the 3D gauge glass model. The nonlinear dynamical
response far below the glass transition  is studied systematically.
A non-Arrhenius type creep motion in the VG phase is observed,
contrarily to the recent studies of the flux line system with strong
pinning. The combined depinning exponent $\beta \delta = 3/2$ is
consistent with those in the 3D random-field Ising model and  the
flux line system with  weak pinning, suggesting the common
universality class in the depining transitions in these systems.

\section{Acknowledgements}

The author acknowledges useful discussions with X. Hu, X. G. Li,
L. H. Tang, and H. H. Wen. The present simulations were partially
performed on SR11000 (HITACHI) at NIMS, Japan. This work was
supported by National Natural Science Foundation of China under
Grant No. 10574107, PNCET and PCSIRT  in University in China,
National Basic Research Program of China (Grant No. 2006CB601003).

\end{document}